\documentclass[preprint2,longabstract]{aa}
\usepackage{epsfig}
\usepackage{graphicx}
\usepackage{amssymb}
\usepackage{natbib}
\def\xte{XTE\,J1810$-$197}

\def\ea {1E\,2259$+$586}

\def\XMM{{\em XMM$-$Newton}}
\def\CXO{{\em Chandra}}
\def\RXTE{{\em R}XTE}
\def\ergscm2{\rm erg\,cm^{-2}\,s^{-1}}
\def\ergs{\rm erg\,s^{-1}}

\begin{document}

\title{From outburst to quiescence: the decay of the transient AXP XTE
J1810$-$197}

\author{F. Bernardini\inst{1,2}
\and G. L. Israel\inst{2}
\and S. Dall'Osso\inst{4,2}
\and L. Stella\inst{2}
\and N. Rea\inst{3}
\and S. Zane\inst{6}
\and R. Turolla\inst{6,7}
\and R. Perna\inst{8}
\and M. Falanga\inst{9}
\and S. Campana\inst{5}
\and D. G\"otz\inst{9}
\and S. Mereghetti\inst{10}
\and A. Tiengo\inst{10}}

\offprints{F. Bernardini: bernardini@oa-roma.inaf.it}
\titlerunning{XTE\,J1810$-$197: from outburst to quiescence}
\authorrunning{Bernardini et al.}

\institute{Universit\`a degli Studi di Roma ``Tor Vergata"
Via Orazio Raimondo 18, I$-$00173 Roma, Italy
\and INAF $-$ Osservatorio Astronomico di Roma, Via Frascati 33, 
I$-$00040 Monteporzio Catone (Roma), Italy.
\and University of Amsterdam, Astronomical Institute ``Anton Pannekoek'', Kruislaan, 403, 1098~SJ, Amsterdam, The Netherlands
\and Dipartmento di Fisica ``Enrico Fermi'', Universit\'a di Pisa,
Largo B. Pontecorvo 3, I$-$56127 Pisa, Italy
\and INAF $-$ Osservatorio Astronomico di Brera, Via Bianchi
	46, I$-$23807 Merate (Lc), Italy
\and Mullard Space Science Laboratory
University College of London
Holmbury St Mary, Dorking, Surrey, RH5 6NT, UK
\and Department of Physics, University of Padova, Via Marzolo 8, I$-$35131 Padova, Italy
\and JILA, Univ. of Colorado, Boulder, CO 80309$-$0440, USA
\and CEA Saclay, DSM/DAPNIA/Service d'Astrophysique, F$-$91191, Gif sur Yvette, France
\and INAF $-$ Istituto di Astrofisica Spaziale e Fisica Cosmica Milano, Via Edoardo Bassini 15, 20133 Milano, Italy}

\date{}

\abstract{}
{\xte\ is the first transient Anomalous X$-$ray Pulsar ever
discovered. Its highly variable X$-$ray flux allowed us to study the timing and spectral emission
properties of a magnetar candidate over a flux range of about two orders of magnitude.}
{We analyzed nine \XMM\ observations of \xte\ collected over a four years baseline 
(September 2003 $-$ September 2007). EPIC PN and MOS data were reduced and
used for detailed timing and spectral analysis. Pulse phase
spectroscopic studies were also carried out for observations with
sufficiently high signal to noise.}
{We find that: (i) a three blackbodies 
model
reproduces the spectral properties of \xte\ over the entire
outburst statistically better than the two blackbodies model previously 
used in the literature, (ii) the
coldest blackbody is consistent with the thermal emission from the whole surface, and has temperature and
radius similar to those inferred from \textit{ROSAT} observations before the outburst
onset, (iii) there is a spectral feature around 1.1\,keV during six consecutive
observations (since March 2005); if due to proton resonant cyclotron scattering,
it would imply a magnetic field of $\sim2\times10^{14}$ G. This is in a very good
agreement with the value from the spin period measurements.}
{} 


\maketitle

\section{Introduction}
Despite isolated neutron stars as a whole are relatively poor X$-$ray emitters,
two small classes of objects stand out for their
widely variable high energy emission, which covers several orders of
magnitude both in intensity and in timescales. These objects are the
Anomalous X$-$ray Pulsars (AXPs; ten objects plus one candidate) and Soft
$\gamma$$-$ray Repeaters (SGRs; 5 objects plus 2 candidates; for a
review see Woods et al. 2006). It is believed that AXPs and SGRs are
linked at some level, owing to their similar timing properties (spin
periods in the 2$-$12\,s range and period derivatives $\dot{P}$ in the
$10^{-13}\div10^{-11}$~s\,s$^{-1}$ range). Both classes have been
proposed to consist of neutron stars whose emission is powered by the decay of
their extremely strong internal magnetic fields ($>10^{15}$\,G; Duncan
\& Thompson 1992, Thompson \& Duncan 1995). Different types of X$-$ray
flux variability are displayed by AXPs. From slow and moderate flux
changes (up to a factor of a few) on timescales of years (virtually
all of the objects of the class), to moderate$-$intense outbursts (flux
variations of a factor up to 10) lasting for 1$-$3 years
(1E\,2259$+$586, and 1E 1048.1$-$5973), to dramatic and intense SGR$-$like
burst activity (fluences of $10^{36}-10^{38}$ ergs) on sub$-$second
timescales (4U\,0142$+$614, XTE\,J1810$-$197, 1E\,2259$+$586 and 1E
1048.1$-$5973; see for a review on the X$-$ray variability see Kaspi et
al. 2007). The first notable recorded case of flux variability was the
2002 bursting/outbursting event detected from 1E\,2259$+$586, in which
a factor of $\sim$10 persistent flux enhancement in an AXP was
followed by the onset of bursting activity during which the source
emitted more than 80 short bursts (Gavriil et al. 2004, Woods et
al. 2004). The timing and spectral properties of the source changed
significantly and attained the pre$-$bursting activity values within a
few days.\\ However, it was only in 2003 that the first transient AXP
(TAXP), namely \xte, was discovered (Ibrahim et al. 2004). This source
was serendipitously detected by the \RXTE\ satellite, and then localized
and studied in greater detail with the \CXO\, and \XMM\ observatories
(Gotthelf et al. 2004, Israel et al. 2004; Rea et al. 2004; Gotthelf
\& Halpern 2005; 2007). It displayed a persistent flux enhancement by
a factor of $>100$ with respect to the quiescent luminosity level of
$\sim 10^{33}\ergs$ (as observed by \textit{ROSAT} and
\textit{Einstein} observatories). Unfortunately, the initial phases of
the outburst were missed and we do not know whether a bursting
phase, similar to that of \ea, occurred also for this source
soon after the onset of the outburst. However, four bursts were
detected by \RXTE\, between September 2003 and April 2004 and
unambiguously associated with \xte\ (Woods et al. 2005). By using
\textit{Very Large Array} (VLA) archival data, Helfand et al. (2005)
discovered a transient radio emission with a flux of
$\sim$$4.5$$\pm$$0.5$\,mJy at 1.4\,GHz at the \CXO\ X$-$ray position of
\xte. Only later, this emission was discovered to be pulsed, highly
polarized and with large flux variability even on very small
timescales (at variance with all known radio pulsars; Camilo et
al. 2006). The VLA data were also used to infer a proper motion of
$13.5\pm1.0$\,mas\,yr$^{-1}$, which, assuming a distance of
$3.5\pm0.5$\,kpc, results in a transverse
velocity of $212\pm35$\,km\,s$^{-1}$ ($1\sigma$ confidence level; Helfand et
al. 2007).\\ Deep IR observations revealed a weak, $K_{\rm s}=20.8$ mag
counterpart, with characteristics similar to those of other AXPs
(Israel et al. 2004). Variability in the IR counterpart of \xte\ was
found (Rea et al.~2004), but it did not correlate with the X$-$ray
emission, contrary to earlier claims (Camilo et al. 2007a; Testa et
al. 2008). It is unclear at present whether the IR variability
correlates with that observed in the radio pulsed emission (Camilo et
al. 2006, 2007a).\\ TAXP are fairly rare objects: a second TAXP was
revealed in 2006 when a candidate AXP, namely CXOU\,J164710.2$-$455216,
displayed a rather intense burst followed by an outburst with a
maximum flux enanchement $>300$, characterized by extreme
changes in both the spectral and timing properties (Muno et al. 2006a,
2006b; Israel et al. 2007a). At variance with \xte,
CXOU\,J164710.2$-$455216 did not show any radio emission so far. The
third TAXP, 1E1547.0$-$5408, was discovered in 2007 when its X$-$ray flux
raised by a factor of $\sim$20 above the quiescent flux. As in the case of
\xte, 1E1547.0$-$5408 was found to be a transient radio
pulsar. Unfortunately the observations missed the outburst onset
(Camilo et al. 2007b; Gelfand \& Gaensler 2007; Halpern et
al. 2008). The three TAXPs above are characterized by a quiescent
state, the timing and spectral properties of which are similar to
those of thousands of other X$-$ray sources present in the
\textit{ROSAT} catalogues: no pulsations (with the exception of
CXOU\,J164710.2$-$455216) and soft X$-$ray spectra well fitted by a
blackbody ($BB$) model with a $kT$ of about 0.1$-$0.2\,keV; again with 
the exception of CXOU\,J164710.2$-$455216, which has $kT\sim 0.5$\,keV 
(Muno et al. 2006b; Skinner et al. 2006). The transient nature of these
three AXPs implies that a relatively large number of members of this
class has not been discovered yet, and suggests that others will
manifest themselves in the future through their outbursts.\\ After
more than four years of data since the outburst onset,
\xte\ provides the first opportunity to study the timing and spectral
evolution of a TAXP as it returns to its quiescent state. Since the first \XMM\ 2003 observations of \xte\ (Gotthelf et al.
Halpern 2004), carried out approximately one year after 
the outburst, it was evident that the source spectrum (two
blackbodies with $kT_{\rm 1}=0.29\pm0.03$\,keV,
$R_{\rm BB1}\sim5.5$\,km, and $kT_{\rm 2}=0.70\pm0.02$\,keV, 
$R_{\rm BB2}\sim1.5$\,km; $L_{\rm X}\sim5\times10^{34}\,\ergs$ in the
0.5$-$10\,keV range for a distance of 3.5\,kpc) was significantly
different from that in quiescence recorded by \textit{ROSAT} in 1992
(one $BB$ with $kT\approx0.18$\,keV and $R_{\rm BB}\approx10$\,km;
extrapolated luminosity in the 0.5$-$10\,keV range of
$L_{\rm X}\sim7\times10^{32}\ergs$; Gotthelf
et al. 2004). Moreover, the source showed 5.54\,s pulsations with a
pulsed fraction of about 45\% during outburst, while only an upper limit
$\sim24\%$ was inferred from the \textit{ROSAT} data (Gotthelf et al. 2004). 

The above properties raise a number of important, still unanswered questions: is the soft
$BB$ component detected by \XMM\ evolving into the quiescent $BB$
component seen by \textit{ROSAT}? What happens to the higher
temperature $BB$ component as the source approaches quiescence? What is
the pulsed fraction level of the source in quiescence (if detectable)? 
Is the quiescent emission revealing the Neutron Star (NS) cooling surface?
Did the outburst lead to a permanent change of the timing/spectral properties
such as the pulsed fraction, the flux and temperature or size of the
quiescent $BB$ component of the source? What is the intensity of
the magnetic field of this source?\\ In this paper, we present a first
attempt to answer the above questions through a detailed study of the timing and spectral
evolution of \xte\ during its outburst decay in 2003$-$2007. In
\S\ref{obs} we report the details of the \XMM\ observations and our
data analysis strategy. Results are presented in \S\ref{results},
while their implications are discussed in \S\ref{diss}.

\section{Observations and Data analysis}
\label{obs}

\xte\, was observed with \XMM\, at nine epochs, the first time for
just $\sim$5\,ks, while the remaining eight observations were deeper,
from $\sim11$\,ks to $\sim60$\,ks (Table \ref{tab:pnvsmos}). The
\XMM\, Observatory (Jansen et al. 2001) includes three
$\sim1500$~cm$^2$ X$-$ray telescopes with an EPIC instrument in each
focus, a Reflecting Grating Spectrometer (RGS; den Herder et al. 2001)
and an Optical Monitor (Mason et al. 2001). Two of the EPIC imaging
spectrometers use MOS CCDs (Turner et al. 2001) and one uses a PN CCD
(Str\"uder et al. 2001). Data have been processed with SAS version
7.1.0, using the updated calibration files (CCF) available in June
2008. Standard data screening criteria are applied in the extraction
of scientific products. We have cleaned all observations from solar
flares by collecting CCD light curves above 10\,keV and applying an
intensity threshold. We also used a time window criterion for removing
solar flare intervals and checked that no significant spectral
differences were present with respect to the intensity threshold
method.\\ During the September 2003 observation, the PN camera was set
in {\tt primary small window} imaging mode with a thin filter (time
resolution=$5.07\times10^{-3}$\,s), while all other observations were
in a {\tt primary large window} imaging mode with a medium filter
(time resultion=$4.76\times10^{-2}$\,s). All observation set$-$ups for
MOS1 and MOS2 cameras were the same, with a time resolution of 0.3 s:
{\tt prime partial window} imaging mode and medium filter (in
September 2003 the MOS1 was set in {\tt prime full window} imaging
mode, in September 2004 the MOS2 was in {\tt Timing uncompressed mode}
and data from this were not reduced). In order to extract more than
90\% of the source counts, we accumulated a one$-$dimensional image and
fitted the 1D photon distribution with a Gaussian. Then, we extracted
the source photons from a circular region of radius 55\arcsec\
($\sim90\%$ of photons) centered at the Gaussian centroid. The
background for the spectral analysis is obtained (within the same PN
or MOS CCD where the source lies) from an annulus region (inner and outer radii
of 65\arcsec\ and 100\arcsec, respectively)
centered at the best source position. In the timing analysis the
background was estimated from a circular region of the same size as
that of the source. All of the EPIC spectra were rebinned before
fitting, in order to have at least 40 counts per bin and prevent
oversampling the energy resolution by more than a factor of
three. Thanks to the time and spectral resolution of the EPIC
instruments\footnote{http://xmm.esac.esa.int/external/xmm\_user\_support/...\\...documentation/uhb\_2.5/node28.html},
we could carry out timing and spectral analysis over the entire set of observations and the Pulse
Phase Spectroscopy (PPS) for the observations with sufficientrly high signal to noise. We
report here the analyses obtained with the PN data and, for
comparison, also the results from the two MOS cameras.
\begin{table*}
\caption{Main observational parameters for the nine \XMM\
datasets. Uncertainties at 1$\sigma$ confidence level are reported. The nine PN
spectra were fitted together according to the prescription discussed in the
text (\ref{feature}); the resulting reduced $\chi^2$ ($\chi^2_{\rm red}$) is 1.09 (1038 d.o.f.), for the 3$BB$$+$edge model.}
\tiny
\begin{center}
\begin{tabular}{cccccccc}
\hline \hline
Epoch & Period & Instrum. (mode)  & Exp.~Time  &  tot ph$-$bck ph   &
$\tau_{\rm max}$ edge& Energy edge & $\chi^2_{\rm red}$\\\\
      & s                    &          &       s         
&                  &
                & keV & \\
\hline \hline
Sep 2003 &5.53928(3)    &EPN$^b$   &5199 & 60136 $-$ 2903      & $<0.17$& 1.10
(fix)& \\
         & &MOS1$^d$ & 7700  & 30761 $-$ 111     & $<0.02 $& 1.10 (fix)& 
1.33\\
         & &MOS2$^e$ & 7800 &  26739 $-$ 145     & $<0.03 $&1.10 (fix)&\\
Mar 2004 &5.53945(1)    &EPN$^c$  & 10730   & 71180 $-$ 3077    & $<0.18 $   &
1.10 (fix)& \\
         &           &MOS1$^e$ & 12000   & 27932 $-$ 396     & $<0.03 $   
&1.10
(fix)& 1.15\\
         &           &MOS2$^e$ & 12200   & 28809 $-$ 366     & $<0.05 $   
&1.10
(fix)&\\
Sep 2004 &5.539599(6)   & EPN$^c$  & 21306    & 89082 $-$ 1574    & $<0.17 
$&1.10
(fix)&  \\
         &            &MOS1$^e$  & 24000    & 35515   $-$ 263   & $<0.06 
$&1.10
(fix)& 1.22\\
         &            &MOS2      & timing mode     &  timing mode            & timing mode    & 
timing mode& timing mode\\
Mar 2005 &5.539825(6)   &EPN$^c$  & 24988  & 54279 $-$ 1760   & $0.12\pm0.03$&
$1.14\pm0.02$ & \\
         & &MOS1$^e$ & 37800  & 26501  $-$ 428  & $0.09\pm0.04$& 
$1.10\pm0.03$ & 1.05   \\
         & &MOS2$^e$ & 37800  & 28004  $-$ 330  & $0.10\pm0.02$& 
$0.98\pm0.08$ & ``   \\
Sep 2005 &5.54004(1)   & EPN$^c$ & 19787  & 21876 $-$ 1311   & $0.26\pm0.04$&
$1.07\pm0.02$ &   \\
         & &MOS1$^e$ & 30000  & 10562  $-$ 344    & $0.2\pm0.1$& 
$1.13\pm0.03$  
 & 1.05 \\
         & &MOS2$^e$ & 18000  & 6645  - 146    & $<0.4$& 1.10 (fix)& `` 
\\
Mar 2006 &5.54022(3)   &EPN$^c$  & 15506  & 12296 $-$ 1197   & $0.17\pm0.05$&
$1.11\pm0.03$&  \\
         & &MOS1$^e$ & 26500  & 6539  $-$ 338    & $0.3\pm0.3$& 
$1.48\pm0.06$&0.95 \\
         & &MOS2$^e$ & 28000  & 7119  $-$ 328    & $0.2\pm0.1$& 
$1.02\pm0.02$& ``  
\\
Sep 2006 &5.54037(1)  &EPN$^c$  & 38505  & 23842 $-$ 2974   & $0.13\pm0.03$&
$1.07\pm0.02$&   \\
         & &MOS$^e$1 & 46800  & 8113  $-$ 552    & $0.20\pm0.05$& 
$1.02\pm0.02$  
&1.47 \\
         & &MOS2$^e$ & 46500  & 8836  $-$ 558    & $0.08\pm0.06$& 
$0.93\pm0.03$& `` 
    \\
Mar 2007 &5.54041(1)  &EPN$^c$  & 37296  & 21903 $-$ 2215   & $0.21\pm0.04 $&
$1.07\pm0.02$ &  \\
         & &MOS1$^e$ & 63000  & 4410  $-$ 1897   & $0.14\pm0.04  $& 
$0.93\pm0.02$
&0.99 \\
         & &MOS2$^e$ & 53000  & 4635  $-$ 522    & $0.15\pm0.05 $&  
$1.10\pm0.02$
& ``\\
Sep 2007 &5.540472(7)  &EPN$^c$  & 59014  & 34386 $-$ 4117   & $0.18\pm0.02$&
$1.04\pm0.06$ &  \\
         & &MOS$^e$1 & 67910  & 11038  $-$ 752    & $0.17\pm0.04$& 
$1.09\pm0.02$ 
 & 1.26\\
         & &MOS2$^e$ & 68785  & 12328  $-$ 768    & $0.10\pm0.03$& 
$1.1\pm0.1$   
& ``  \\
\hline \hline
\end{tabular}
\label{tab:pnvsmos}
\end{center}
\end{table*}

\begin{figure*}
\begin{center}
\includegraphics[angle=+90,scale=.635]{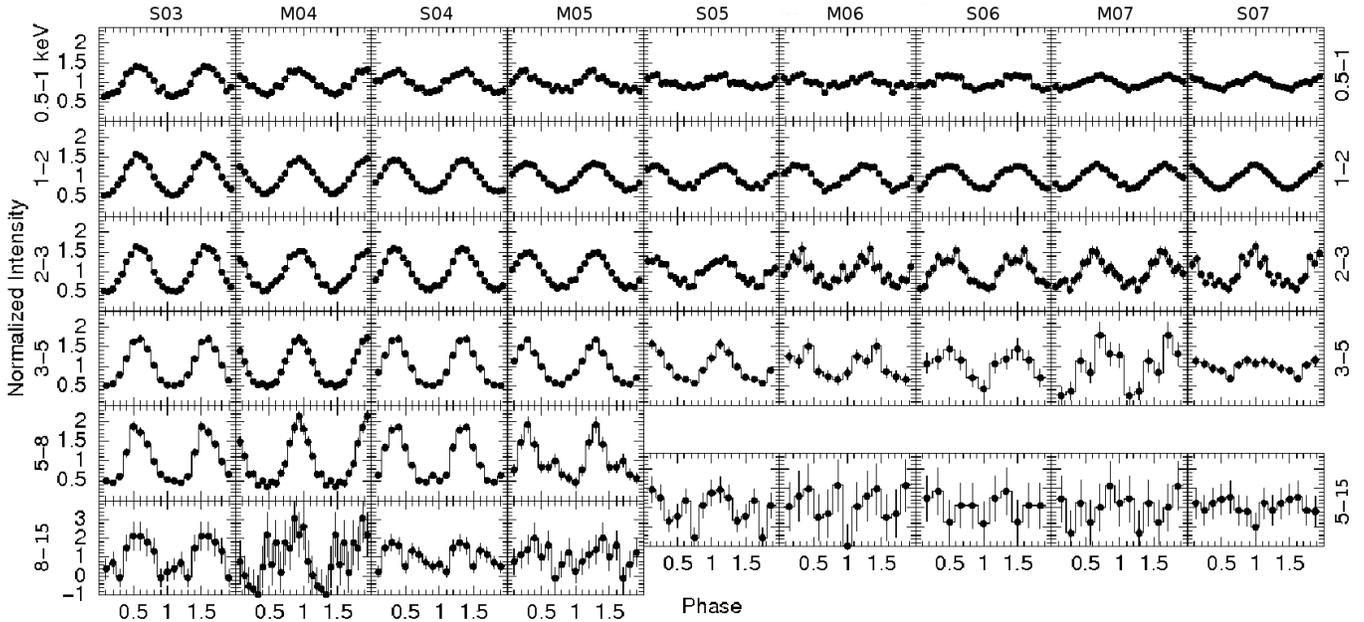}
\caption{\xte\ PN background subtracted light curves folded at the best period (see Table\,\ref{tab:pnvsmos}) for each of the nine \XMM\ observations carried out
between September 2003 (S03) and September 2007 (S07), and for different energy intervals: 0.5$-$1\,keV, 1$-$2\,keV,
2$-$3\,keV, 3$-$5\,keV, 5$-$8\,keV, and 8$-$15\,keV. The last two energy intervals
have been merged together since the September 2005 (S05) pointing in order to 
improve the statistics.} 
\label{figure:propuls}
\end{center}
\end{figure*}

\section{Results}
\label{results}

\subsection{Timing analysis}
\label{sect:analtemp}

The source event arrival time of each observation, in the 0.5$-$15\,keV
energy range, were converted into barycentric dynamical times (BDT) by
means of the SAS tool {\tt barycen} and the ($\sim$1$''$ accurate)
source position provided by Helfand et al. (2007). Given the complex
time evolution of the period and its derivatives as derived by radio
observations (Camilo et al. 2007c), we measured only the local spin period 
at each single \XMM\ pointing by means of a phase fitting
technique (events in the 0.5$-$10\,keV energy range were used; see
e.g. Dall'Osso et al. 2003 for details on the technique). Different period
measurements are independent and not phase$-$connected. Folding each lightcurve at its
measured spin period we obtained the pulse profile and found that it
remained single peaked in all observations (Figure
\ref{figure:propuls}). In order to estimate the dipole field strength
of this source we refer to the phase$-$coherent measurements of $\nu$, $\dot{\nu}$, and
$\ddot{\nu}$ obtained by Camilo et al. (2007c).
These authors measured fast variations of $\dot{\nu}$ that did not
allow them to provide a unique value for the magnetic field strength. The frequency derivative
was found to change continuously over 300 days of monitoring from 
$\sim-3.4\times10^{-13}\,\rm s^{-2}$ to
$\sim-1.4\times10^{-13}\,\rm s^{-2}$. Accordingly, we consider the
secular spindown trend as bracketed by these limits and derive a
corresponding range of values for the magnetic field
$1.6\times10^{14}\,{\rm G}\leq B_{\rm dip}\leq2.8\times10^{14}\,\rm G$
through the standard dipole formula.

\subsubsection{Pulsed fraction}
\begin{figure*}[t!]
\vspace{1.0 cm}
\includegraphics[angle=-90, scale=0.8]{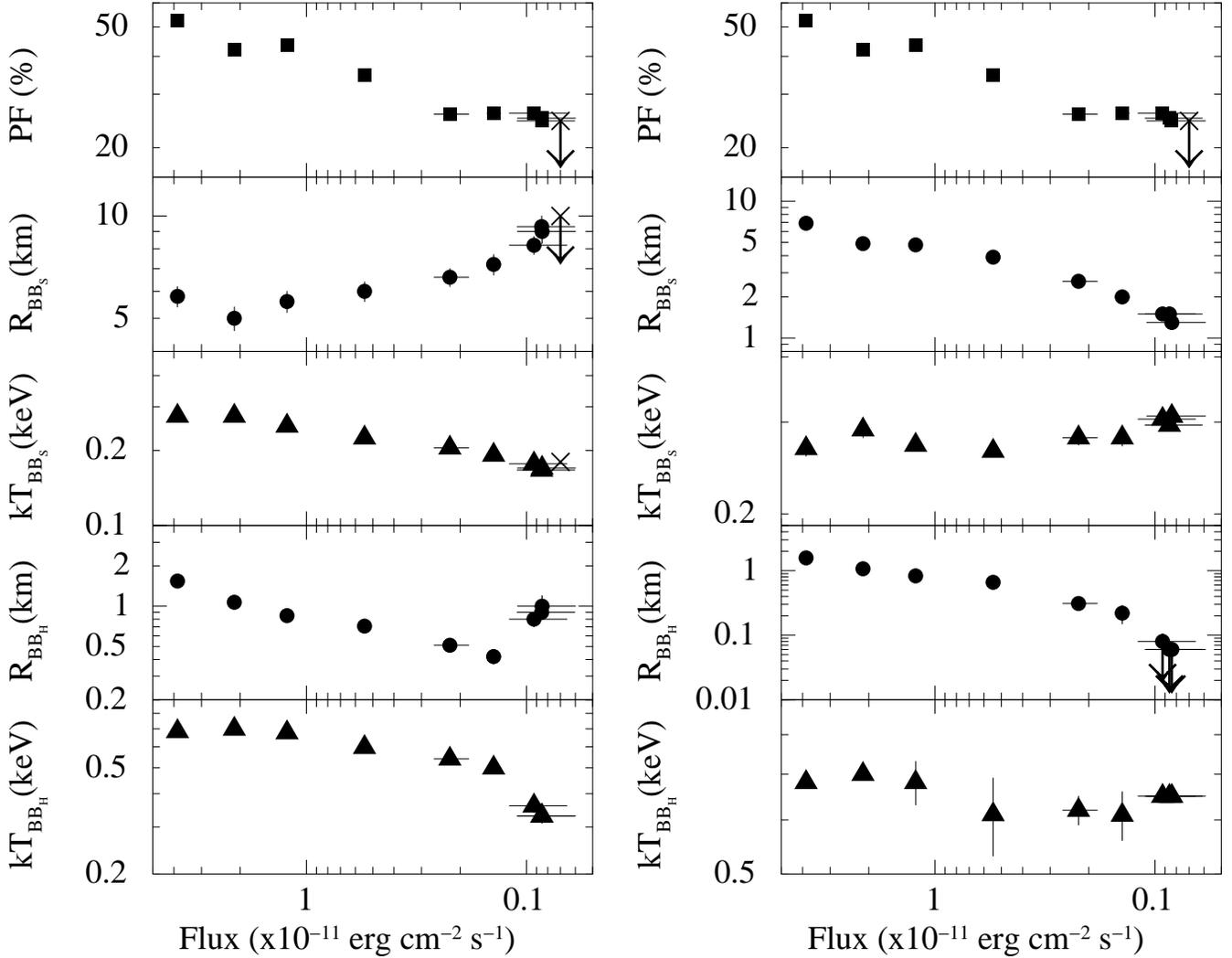}
\caption{Evolution of the spectral parameters for the
2$BB$ (left panel) and 3$BB$ (right panel) models together with the pulsed fraction (PF)
as a function of the 0.6$-$10\,keV flux. The cross in the left panel (first, second, and third row), and in the right panel (first row) marks the \textit{ROSAT} data.}
\label{fig:XTE}
\end{figure*}

Given the smooth and nearly sinusoidal pulse shape, we
could determine with reasonable accuracy the Pulsed Fraction (PF) of
the signal (defined here as: $\rm PF=(A_{max}-A_{
min})/(A_{max}+A_{min})$, where $\rm A_{max}$ and $\rm A_{min}$ are the
maximum and minimum of the sinusoidal modulation). Between September
2003 and September 2007, the PF decreased by a factor of about two
(between $\sim50\%$ and $\sim25\%$) in the $0.5-10$ keV energy
interval (Figure \ref{fig:XTE}). In particular, since March 2005, the
PF in the $0.1-2.5$ keV band reached $\sim(25\pm1)\%$ (here and troughout this paper uncertainties are given
at $1\sigma$ confidence level, where not stated otherwise); this value is
close to the upper limit ($\sim24\%$) inferred from the 1992$-$1993 \textit{ROSAT}
pointings during the quiescent phase of the source (Gotthelf et
al. 2004).\\ 
Moreover, the PF decreases as a function of time in the same energy band and
increases as a function of energy within the same observation, as shown in Table
\ref{tab:FPvsE}. Between 8 and 15\,keV the pulsed fraction is
consistent with $100\%$ ($3\sigma$ confidence level). However, the
relatively poor statistics above 10\,keV prevented a detailed study of
the spectral properties of this high energy component (see also
Section\,\ref{section:spec}).\\
\begin{table*}
\caption{Pulsed fraction (PF) in different energy intervals 
vs time, and $\rm PF^{\rm total}_{\rm 0.5\div15 }$ keV. Errors are 
reported at $1\sigma$ confidence
level.}
\tiny 
\begin{center}
\begin{tabular}{cccccccc} 
\hline \hline
\\
Epoch &$\rm PF_{0.5\div1\, keV}$ & $\rm PF_{1\div2\, keV}$ & $\rm PF_{2\div3\, keV}$ &
$\rm PF_{3\div5\, keV}$  & $\rm PF_{5\div8\, keV}$  & $\rm PF_{8\div15\, keV}$ & $\rm PF^{total}_{0.5\div15\, keV}$\\
    &\% &\% &\% &\% & \%& \% &\%\\
\hline \hline
Sep 2003& $ 35  \pm  2 $  &  $49.3 \pm 0.5$ & $57.0 \pm 0.8$ &  $58.7\pm  0.9$  
& $63\pm   3$ &$ 100  \pm 20$ &$52.4\pm0.4$\\
Mar 2004& $ 25  \pm  1 $  &  $37.2 \pm 0.5$ & $42.1 \pm 0.8$ &  $51\pm 1$  
& $66\pm  3$ &$ 50 \pm 25$ &$42.0\pm0.4$\\
Sep 2004& $ 27  \pm  1 $  &  $40.2 \pm 0.7$ & $49.6 \pm 0.7$&  $57.6\pm  0.9$  
& $63\pm   2$ &$ 50  \pm 12$ &$43.5\pm0.3$\\
Mar 2005& $ 20  \pm  1 $  &  $33.1 \pm 0.6$ & $44   \pm 1$ &  $51  \pm  1$  
& $49\pm   5$ &$ 100  \pm 40$ &$34.8\pm0.4 $\\
Sep 2005& $ 13  \pm  2 $  &  $27.2 \pm 0.9$ & $32   \pm 2$ &  $30.0  \pm  0.9$
& $60\pm  11$ &$ 60  \pm 30$ &$25.8\pm0.7$\\
Mar 2006& $ 13  \pm  2 $  &  $30   \pm 1$ & $32   \pm 3$ &  $26  \pm  6$  
& $70\pm  29$ &$ 20  \pm 48$ &$26\pm1$\\
Sep 2006& $ 16  \pm  1 $  &  $29.3 \pm 0.9$ & $42   \pm 3$ &  $90  \pm 72$  
& $90\pm  72$ &$ 40  \pm 41$ &$26\pm0.7$\\
Mar 2007& $ 15  \pm  2 $  &  $32   \pm 1$ & $36   \pm 3$ &  $60  \pm  10$ 
& $30\pm  33$ &$ 40  \pm 10$ &$25\pm1$\\
Sep 2007& $ 15  \pm  1 $  &  $28.1 \pm 0.7$ & $41   \pm 3$ &  $50  \pm 10$  
& $40\pm  50$ &$ 80  \pm 25$ &$24.5\pm0.6$\\
\hline \hline
\end{tabular}
\label{tab:FPvsE}
\end{center}
\end{table*}
\subsection{Spectral analysis}
\label{section:spec}
In the following we describe a detailed spectral analysis of our \XMM\ dataset which includes the outburst evolution
down to its almost complete decay. The outburst spectrum in its brightest phase had already been analyzed in the literature with a 
two blackbodies spectral model (2$BB$, Gotthelf et al. 2004, Gotthelf \& Halpern 2005). On 
the other hand the quiescent emission from the source recorded by \textit{ROSAT} was consistent with a different, single $BB$. Starting from this, our strategy was then twofold: first we tried to apply the 2$BB$ model to the whole \XMM\ dataset to check whether one of the two components evolved smoothly to the quiescent one. Then we tested an alternative possibility, namely that the quiescent component was independent and always present, the spectrum of the outburst being superimposed on it; this led us to consider a phenomenological model including three different thermal components (3$BB$ model). In this scenario, as the outburst flux decays its spectral components progressively fade away eventually revealing the underlying quiescent emission. As such, the quiescent component could be tentatively identified with the thermal emission from the whole NS surface.\\

\subsubsection{Thermal components}
\label{sub:2bb-3bb}
Following Gotthelf et al. (2004)
and Gotthelf \& Halpern (2005), we applied the 2$BB$ spectral fit to the
fading phases of \xte\ until September 2007, when the source flux was
$\sim$1.2 times higher than the pre$-$outburst level ($\chi^2_{red}\sim1.26$ for 1058 d.o.f., which is at $6\sigma_{\rm \chi^{2}}$
from the expectation value\footnote{$\sigma^{2}_{\rm \chi^{2}}= 2
  dof\Rightarrow\sigma_{\rm \chi^{2}}=\rm \sqrt{2dof}$,
  $\frac{(\chi^{2}-dof)}{\sigma_{\rm \chi^{2}}}= x[\sigma_{\rm \chi^2}]$,
  where x is the distance from the expectation value of $\chi^{2}$ in
  unit of $\sigma_{\rm \chi^2}$.};
$N\rm_H$=$(0.60\pm0.01)\times10^{22}$\,cm$^{-2}$; see also Table
\ref{tab:ragtempPBB}). The 2$BB$ model (see also Perna \& Gotthelf
2008 for a detailed study) corresponds to a scenario in
which one of the two $BB$ naturally evolves into the single $BB$
spectrum detected by \textit{ROSAT} in the quiescent state while the
other (hard) $BB$ just fades away. The results of this approach showed
that, while the cold $BB$ component smoothly approaches the quiescent one
(see Figure \ref{fig:XTE}, left panel, 2$^{nd}$ and 3$^{rd}$ plot), a
number of ambiguities arise. The radius of the hot $BB$ does not decrease
monotonically with flux (time): after 2.5 years of smooth decrease it starts increasing in September
2006 (left panel, 4$^{th}$ plot). At the same epoch its temperature drops
rapidly reaching a value comparable to that of the cold $BB$ in the first part of the outburst (Figure \ref{fig:XTE}, left panel, 5$^{th}$
plot). Moreover, neither spectral component is able to account
for the flattening of the pulsed fraction at the 25\%
level, (Figure \ref{fig:XTE}, left panel, 1$^{st}$ plot).\\
These findings suggest that the observed emission might come from a more complex structure than a simple two$-$component model and that we might be seeing different parts of the whole structure as the flux decreases. With this scenario in mind we repeated the spectral analysis using the 3$BB$ model discussed in the previous subsection.

\begin{figure*}
\begin{center}
\includegraphics[angle=-90,scale=.70]{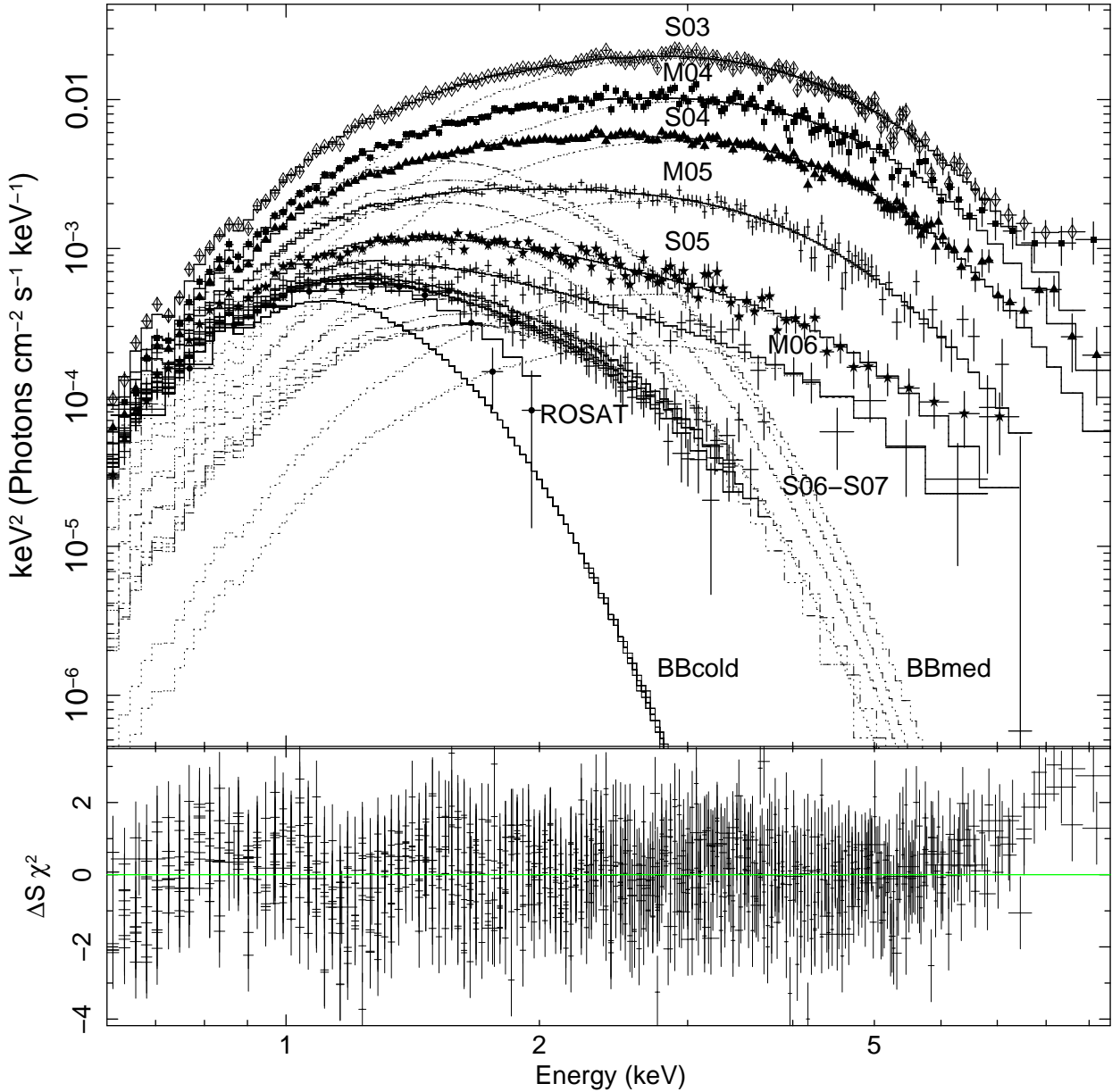}
\caption{Spectral evolution in the nine \XMM\ observations, and the
\textit{ROSAT} spectrum in quiescence. Fits use the 3$BB$ model (upper panel). Model residuals are shown in the
lower panel. The soft component, constant in time, is marked with $BB_{\rm cold}$; the medium$-$temperature evolving blackbody component is marked with $BB_{\rm med}$.
There is a sharp drop in the flux above 2.5 keV in the last three
spectra (compare M06 with S06, M07 and S07) which corresponds to the
disappearance of the hot component $BB_{\rm hot}$. M=March,
S=September; 03=2003, 04=2004, 05=2005, 06=2006, and 07=2007.}.\\
\label{figure:spettPBBBrosat}
\end{center}
\end{figure*}

For the first six observations (during which the total flux is significantly higher than the
pre$-$outburst one), all parameters of the 3$BB$ model were left free to vary 
except for $N\rm_H$ 
which was constrained to be the same in all observations. We found that it is always possible
to fit the first 6 data sets (September 2003$-$March 2006) with a 3$BB$
model without forcing the spectral parameters (3$BB$: $\chi^2_{\rm red}=1.1$, 812 d.o.f.; 2$BB$: $\chi^2_{\rm red}=1.21$, 824 d.o.f., F$-$test probability $\simeq10^{-11}\sim7\sigma$).
The extra $BB$ has a characteristic temperature
$kT\sim0.14$ keV which is constant in time, but whose radius could not
be well constrained ($R<100$ km). Under the
hypothesis that the latter component originates from the whole NS surface,
we can consider it constant through the whole
outburst. Correspondingly, we left free to
vary the temperature and radius of this additional $BB$, but forced both parameters to maintain 
the same value in all spectra.\\ 
We then applied the 3$BB$ model to all of the 9 \XMM\ observations.
The addition of the extra $BB$
component gave a better fit as compared with the 2$BB$ model 
($\chi^2\sim1250$, $\chi^2_{\rm red}\sim$1.18 for 1056 d.o.f., $N\rm_H$=$(0.72\pm0.02)\times10^{22}$\,cm$^{-2}$);
an F$-$test gives a 7.3$\sigma$ significance for the
inclusion of the additional spectral component. Notably, the overall fit gave
parameters for the coldest $BB$, $kT_{\rm cold}=0.144\pm0.003 $ keV;
$R_{\rm cold}=17.9\pm^{1.9}_{1.5}$ km, $F^{\rm 0.1-2.5\,
kev}_{\rm X}=(4.5\pm0.5)\times10^{-13}$ erg cm$^{-2}$ s$^{-1}$ which are
very close to those inferred in quiescence with \textit{ROSAT}
($kT=$0.18$\pm$0.02 keV and $R\sim10$ km,
$F^{\rm 0.1-2.5}_{\rm X}\sim5.4\times10^{-13}$ erg cm$^{-2}$ s$^{-1}$). Even
more interesting, the two hotter $BB$ components maintained a nearly
constant temperature as the source flux decayed in time (see
Figure\,\ref{fig:XTE}, right panel, 3$^{rd}$ ad 5$^{th}$ plots). Their
radius appears to be the only variable parameter during the decaying
phase of the outburst (Figure \ref{fig:XTE}, right panel, 2$^{nd}$ and
4$^{th}$ plots).\\ Starting from September 2006 the spectrum could be well fitted
by a simple 2$BB$ model. The hottest component was not needed anymore and we could set a
$3\sigma$ upper limit on its flux of $\sim8.7\times10^{-14}\,\ergscm2$.
The above mentioned flattening of the pulsed fraction at this epoch could be accounted
for quite naturally by the disappearance of this hot component (Figure \ref{figure:spettPBBBrosat}).\\
\begin{table}[h]
\caption{Temperature (keV) and radius (km) evolution with time of $BB_{\rm cold}$ 
and $BB_{\rm hot}$ in the 2$BB$ model. Uncertainties are at $1\sigma$ confidence level (68\%).}
\tiny
\begin{center}
\begin{tabular}{ccccc}
\hline 
\hline
Epoch & $kT_{\rm cold}$  & $R_{\rm cold}$  & $kT_{\rm hot}$ & $R_{\rm hot}$\\
    & keV  & km  &   keV &  km     \\
\hline 
\hline 
Sep 03 & $0.275\pm0.009$  &$5.8\pm0.4$ &$0.685\pm0.005$ &$1.54\pm0.03$\\ 
Mar 04 & $0.275\pm0.009$  &$5.0\pm0.4$ &$0.699\pm0.009$ &$1.07\pm0.03$\\
Sep 04 & $0.251\pm0.005$  &$5.6\pm0.4$ &$0.677\pm0.005$ &$0.85\pm0.02$\\
Mar 05 & $0.225\pm0.004$  &$6.0\pm0.4$ &$0.597\pm0.006$ &$0.71\pm0.02$\\
Sep 05 & $0.205\pm0.004$  &$6.6\pm0.4$ &$0.54\pm0.01$   &$0.51\pm0.04$\\
Mar 06 & $0.192\pm0.004$  &$7.2\pm0.5$ &$0.50\pm0.02$   &$0.42\pm0.05$\\
Sep 06 & $0.177\pm0.004$  &$8.2\pm0.5$ &$0.36\pm0.02$   &$0.8 \pm0.1$\\
Mar 07 & $0.170\pm0.005$  &$9.0\pm0.7$ &$0.33\pm0.02$   &$1.0 \pm0.2$\\
Sep 07 & $0.167\pm0.004$  &$9.3\pm0.7$ &$0.33\pm0.01$   &$0.9 \pm0.1$\\ 
\hline 
\hline
\end{tabular}
\label{tab:ragtempPBB}
\end{center}
\end{table}
\noindent
Given the stability of both the flux and spectrum over the last three
observations (less than 8\% variation in flux in the $0.6-10$\,keV
energy band; September 2006 $-$ 2007), we merged the source photon lists
in order to obtain a higher S/N spectrum (note that the calibration of
the PN instrument has proved to be also very stable; Krisch et
al. 2005). The analysis of the merged spectrum significantly improved the determination of the spectral parameters as compared 
to each single spectrum (see Table \ref{tab:3bbtot}). Furthermore, the hotter $BB$ component remained statistically non$-$significant also in the merged spectrum. We could thus obtain a more accurate ($3\sigma$) upper limit on its flux: $F_{\rm BBhot}<4.5\times10^{-14}\ergscm2$ in the 0.6$-$10 keV energy band.\\ 
We also attempted to estimate the pulsed fraction of the quiescent emission, which we tentatively attributed to the NS surface,
taking into account the last three observations only, so that the hotter $BB$ was absent.
We note that this is only possible in our scenario since in the 2$BB$ model the softer component
is still evolving towards quiescence.\\
We express the PF in the 0.1$-$1 keV band as: PF$_{(0.1-1\,\rm keV)}=\alpha\rm
F_{cold}+ \beta \rm F_{med}$, where F$_{\rm cold}\sim0.9$ and F$_{\rm med}\sim0.1$ represent the
relative contributions of the two spectral components to the total flux in the 0.1$-$1 keV band. Correspondingly, $\alpha$ and $\beta$ represent their PFs. The value of $\beta$ is obtained from the PF in the
2.8$-$4.1 keV energy range, where $BB_{\rm cold}$ is negligible, and turns
out to be $\simeq17\%$. The PF of the cold $BB$, $\alpha$, is thus completely determined by
the measured value of PF$_{(0.1-1\,\rm keV)}$. We obtain $\alpha=10\pm1\%$, a prediction that can be checked once the source will return to the quiescent state.\\
\begin{table}
\caption{Temperature, radius and observed flux (0.6$-$10
keV) evolution with time of $BB$ medium (med) and hot $BB$ (hot) in the 3$BB$ model. $1\sigma$
uncertainties are reported. Upper limits are inferred at
$3\sigma$ confidence level.: $kT_{\rm cold}=0.144\pm0.003 $ keV; $R_{\rm cold}=17.9\pm^{1.9}_{1.5}$ km,
$N_{\rm H}=(0.72\pm0.02)\times10^{22}\,\rm cm^{-2}$ are constant throughout the outburst.}
\begin{center}
\scriptsize
\begin{tabular}{cccc}
\hline\hline
Epoch & $kT_{\rm med}$ & $R_{\rm med}$ & $F_{\rm med}$\\ 
   & keV & km & erg$\,\rm s^{-1} cm^{-2}$\\
\hline 
Sep 03& $0.267\pm0.009 $&$6.9\pm0.6  $&$(5\pm1)\times10^{-12}  $\\
Mar 04& $0.29\pm0.01   $&$4.9\pm0.4  $&$(4\pm1)\times10^{-12}  $\\
Sep 04& $0.271\pm0.006 $&$4.8\pm0.3 $&$(2.6\pm0.5)\times10^{-12} $\\
Mar 05& $0.264\pm0.007 $&$3.9\pm0.3 $&$(1.5\pm0.3)\times10^{-12} $\\
Sep 05& $0.280\pm0.009 $&$2.6\pm0.2 $&$(9\pm2)\times10^{-13} $\\
Mar 06& $0.28\pm0.01   $&$2.0\pm0.2 $&$(6\pm2)\times10^{-13} $\\
Sep 06& $0.304\pm0.006 $&$1.5\pm0.1 $&$(5\pm1)\times10^{-13} $\\
Mar 07& $0.296\pm0.006 $&$1.5\pm0.2 $&$(4\pm1)\times10^{-13} $\\
Sep 07& $0.308\pm0.006 $&$1.3\pm0.1 $&$(4\pm1)\times10^{-13} $\\
Sep 06-07& $0.301\pm0.003$ & $1.42\pm0.03$& $(4.1\pm0.2)\times10^{-13}$\\
\hline 
Epoch  & $kT_{\rm hot}$ & $R_{\rm hot}$ & $F_{\rm hot}$\\
   &  keV & km & erg$\,\rm s^{-1} cm^{-2}$\\
\hline 
Sep 03&$0.681\pm 0.005   $  &$1.58  \pm 0.04  $ &$(3.3\pm0.2)\times10^{-11}   $ \\
Mar 04&$0.70\pm 0.01    $ &$1.07  \pm 0.04  $ &$(1.7\pm0.2)\times10^{-11}     $\\
Sep 04&$0.68\pm 0.05  $   &$0.83  \pm 0.02  $ &$(9.2\pm0.7)\times10^{-12}   $\\
Mar 05&$0.61\pm 0.08   $  &$0.66  \pm 0.03  $ &$(3.5\pm0.5)\times10^{-12}   $\\
Sep 05&$0.62\pm 0.03    $ &$0.31  \pm 0.05  $ &($8\pm4)\times10^{-13}   $\\
Mar 06&$0.61\pm 0.05    $ &$0.22  \pm 0.07 $ &$(4\pm3)\times10^{-13}   $\\
Sep 06&$0.65  $ $^a$  &$<0.08 $              &$<8.7\times10^{-14}$ \\
Mar 07&$0.65 $ $^a$   &$<0.06 $              &$<6.1 \times10^{-14}$ \\
Sep 07&$0.65 $ $^a$  &$<0.06$                &$<5.4\times10^{-14} $\\
Sep 06-07&$0.65 $ $^a$  &  $<0.05$          & $<4.5 \times10^{-14} $\\
\hline \hline 
\end{tabular}
\label{tab:3bbtot}
$^a$~Fixed to the average of the earliest measurements. 
\end{center}
\end{table}
\subsubsection{The power$-$law component}

By adopting the 3$BB$ model, we further study the possible presence of
additional features in the \XMM\ spectra. In particular, during the
first three \XMM\ observations (2003$-$2004), the spectral fit residuals 
suggest the presence of an additional 
hard component above 7$-$8\,keV (3.2$\sigma$ confidence level) which we were
not able to characterize due to poor statistics in this band. We can only
speculate that it might be related to a hard
power$-$law$-$like tail ($\Gamma\sim1.5$), likely of magnetospheric origin. A similar component has been detected in
other AXPs (Kuiper et al. 2004; Kuiper et al. 2006) and extends up to 200 keV, at least (G\"otz et al. 2006). Given the marginal significance of this component we do not attempt to draw any firm conclusion.

\subsubsection{Narrow spectral feature search}
\label{feature}
\begin{figure}
\begin{center}
\includegraphics[angle=-90,scale=.50]{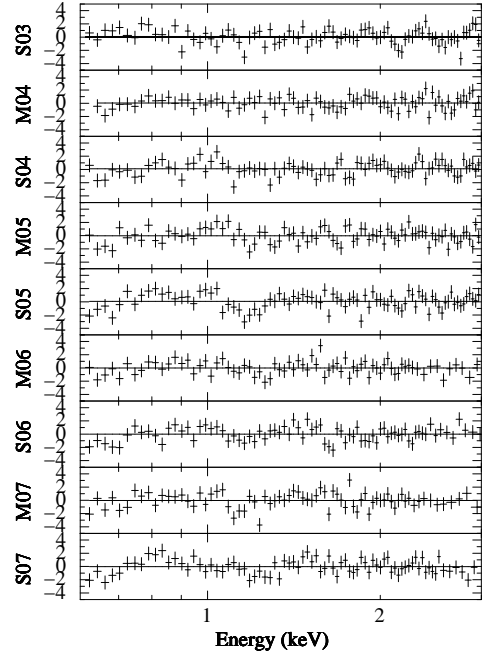}
\caption{$\Delta\chi^2$ residuals of 3$BB$ model. From March 2005 (M05)
onwards is likely present a feature (edge) around $\sim1.1$ keV.}
\label{figure:deltachi}
\end{center}
\end{figure}

\begin{figure}
\begin{center}
\includegraphics[angle=-90,scale=.35]{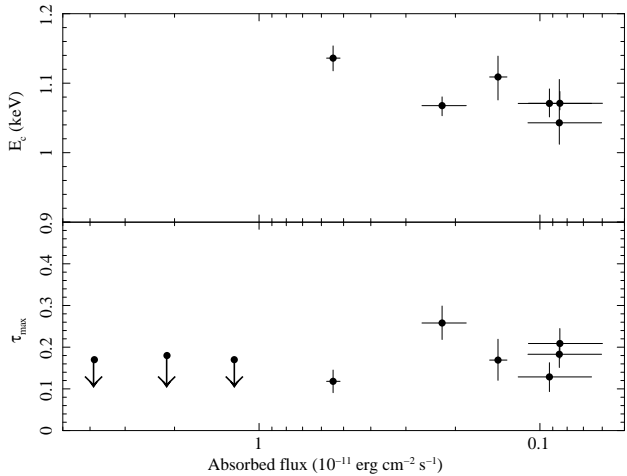}
\caption{Parameters of the $\sim$1.1 keV edge (energy ($E_{\rm c}$) and $\tau_{\rm max}$) vs 
the observed flux. For
the first three observations $\tau_{\rm max}$ is an upper limit.}
\label{figure:evoluzEedgtau}
\end{center}
\end{figure}

Starting from the 4th observation (March 2005) we note the presence of excess residuals in the data with respect to the
3$BB$ model, at around 1.1\,keV (Figure \ref{figure:deltachi}). We tried to account for this by including an absorption edge or a Gaussian line in the model. The value of the former is 1 if $E\leq E_{\rm c}$ and ${\rm exp}[-\tau_{\rm max}(E/{E_{\rm c}})^{-3}]$ if $ E\geq E_{\rm c}$ where $E_{\rm c}$ is the threshold energy and \textit{$\tau_{\rm \max}$} the absorption depth at the
threshold.\\
The results of the new spectral model, 3$BB$ plus edge, are consistent with
what was obtained with the 3$BB$ model (to within the uncertainties):  
$N_{\rm H}=(0.73\pm0.02)\times10^{22}$ cm$^{-2}$, $kT_{\rm cold}=0.153\pm0.005$ keV and
$R_{\rm cold}=15.4\pm1.8$\,km ($\chi^{2}=1140$ with d.o.f.$=1038$). 
The energy threshold ($\sim1.1$ keV) and $\tau_{\rm max}$ ($\sim0.2$) appear
to be constant through the latest six observations (Table \ref{tab:pnvsmos}
and Figure \ref{figure:evoluzEedgtau}).\\
This new model has $\chi^{2}_{\rm red}=1.09$, which is at 2.2$\sigma_{\rm \chi^{2}}$ from the expectation value. 
To obtain an estimate of the significance of the edge component we proceeded as follows:\\
We obtained, for each single spectrum, the width of the feature ($\sigma$) using a Gaussian profile and defined the width at the base of the Gaussian ($\Delta E_{i}$) to be $3\sigma$. We assumed that the width of the feature is independent of the model used to estimate it. Then we calculated the ratio between the whole spectral range of our data, $\Delta W_{\rm i}$, and the Gaussian width $\Delta E_{\rm i}$ (number of trials). Finally, in order to obtain the total probability of the null hypothesis (no line present), we multiplied the probability level ($P_{\rm i}$) attributed by an F$-$Test to the inclusion of the Gaussian by the number of trials on each spectrum 
($\Delta W_{\rm i}/\Delta E_{\rm i}$) and by the total number of observations (9). The total probability can thus be expressed as:
$9\times\Pi_{\rm i}P_{\rm i}\Delta W_{\rm i}/\Delta E_{\rm i}$, which gave in our case a significance for the edge component at the 
$\simeq 6.5\sigma$ level.\\
As a further check, we estimated the line significance by running a
Monte Carlo simulation of $10^5$ spectra with only the continuum model present (as described
in more detail in Rea et al. 2005, 2007). Spectral parameters of the continuum
were allowed to vary within 3 sigma from their best fit values and we used the same number of photons of the 4th
observation (March 2005). We then counted how many edges, at any energy between
0.5-10keV, with $\tau>0.2$ have been significantly detected in the
generated spectra just due to statistical fluctuations. We found 12
spectra over $10^5$ spectra presenting such a feature, thus leading to an estimated significance level of $\sim4.1\sigma$ for our 1.1\,keV edge. However, so far we did not consider that the feature has been detected
in several spectra rather than only in the 4th observation. To this
aim we simulated $10^5$ spectra for each of the 6 observations showing
the $\sim$1.1 edge, using the best fit spectral parameters and the corresponding number of
photons for each observation. However, in these simulations we only
considered the energy band 0.5-4\,keV in which the spectral
variability was not so large among the 6 observations. This reduces to negligible levels 
any possible systematic error in the probability calculation due to the spectral variability 
of the source in connection with instrumental response matrices. 
We estimated the significance of the edge in each observation as described above, then
combined them to obtain a total significance of 5.1$\sigma$ for the presence of the line.

By using a Gaussian profile to fit the feature, we obtained results similar
to those of the edge component, the mean energy of the feature
being $<E>\sim1.15$ keV, $<\sigma>\sim0.13$ keV, while the average
equivalent width of the line was $\sim35$ eV ($\chi^2_{\rm red}$=1.09,
1029 d.o.f).\\ The use of different chemical abundances for the
interstellar medium (ISM), vphabs model in $XSPEC$, does not
produce significant changes in the parameter values or confidence
level of the feature, which thus does not seem to depend on the ISM
composition.\\
In order to test the possible instrumental nature of the feature we
also used the source photons collected by the MOS1 and MOS2 cameras. As in the case of
the PN data, we carried out a spectral analysis by using the 3$BB$ model to account
for the continuum spectral component. Table \ref{tab:pnvsmos} summarizes the
results of this test. Starting from the March 2005 observation, the edge component is always detected in all three cameras
except for the September 2005 observations MOS2 data, where only an upper limit could be obtained. In the latter case the inferred upper limit is consistent with the values inferred from all other spectra. This
finding further supports the interpretation of the edge as intrinsic to the source.

To further check our results we analyzed the only $CHANDRA$ public observation of \xte\ made with $ACIS$, during March 2006 ($\sim30$ ks). We used $CIAO\,4.0$ software, a standard reduction procedure, and the last calibration files availeble (3.5.0, October 2008) for the $CHANDRA$ data analysis. This gives fully consistent results with those obtained with \XMM: $E_{\rm c}=1.05\pm0.3$, $\tau_{\rm max}=0.38\pm0.9$. The significance for the inclusion of this component, determined using the procedure previously exposed, is $\sim4.5\sigma$. This provides further confirmation for the presence of this feature in the continuum of the source. Therefore in the following we consider the 3$BB$$+$edge as our best spectral model. 
\subsubsection{Other models}
\label{sub:compton}
An alternative possibility to model the data is by considering the
effect of resonant Compton scattering (RCS) in the magnetosphere
(Thompson, Lyutikov \& Kulkarni 2002). In this scenario, photons
emitted by the star surface, at the temperature of $\sim0.16$ keV, are
upscattered by energetic electrons and/or positrons in the
magnetosphere. Therefore, the increase in X$-$ray flux during the
outburst would not be due (only) to the appearance of (hotter) regions
with enhanced emission, but to a shift in energy of upscattered
photons.\\ We have performed some tests with a thermal Comptonization
model readily available in $XSPEC$ ($CompTT$, Titarchuk 1994). Although based
on completely different physical assumptions with respect to RCS, this may
at least be used to assess whether the observed spectra can be modelled in
terms of Comptonization. We fitted together all the 9 spectra 
assuming, as a first approximation, that the plasma temperature is the
same at all epochs. A best fit is obtained with an electron temperature
of $kT_{\rm e}\sim0.8$ keV, a constant (within uncertainties)
temperature for the seed photons of $kT_{\rm seed}\sim0.16$ keV and a
plasma optical depth ($\tau_{\rm p}$) decreasing with time from $\sim32$
to $\sim9$. The $\chi^{2}_{red}$, however, is worse than that of the 3$BB$
model, namely $\chi^{2}_{\rm red}=1.3$ for 1066\,d.o.f. (this value is at
6.7$\sigma_{\rm \chi^{2}}$ from the expectation value). We note also that a
scenario in which scattering is (nearly) isotropic and the
Comptonizing medium uniformly covers the star surface is hardly
compatible with the observed characteristics of the pulsed emission. Indeed,
the (relatively) small pulsed fraction of the thermal
component would be further washed away by scattering at higher
energies.\\ Also in this case a feature in the spectrum around 1.1 keV seems to be present. By fitting this feature with an edge component, like in the case of the 3$BB$ model, we obtain a significance level of $\sim6.5\sigma$. Therefore, this feature seems to be independent on the model used for the underlying spectral continuum.

\subsection{Pulse Phase Spectroscopy}

In order to understand the role of each spectral parameter in 
producing the observed $0.6\div10$ keV flux variation with pulse phase, we carried out a pulse phase resolved
spectroscopic analysis of the \XMM\ observations with sufficiently high S/N.
The spectra of the first three observations (September 2003$-$September 2004) were considered and 
divided into 10 phase intervals, in order to rely upon a sufficiently large number of photons.
The spectrum in each phase interval was modelled with 3$BB$s fixing the temperature and radius of
$BB_{\rm cold}$ and $N_{\rm H}$ at the average values obtained from the previous
analysis without the inclusion of the edge ($kT_{\rm cold}=0.144\pm0.003 $ keV; $R_{\rm cold}=17.9\pm^{1.9}_{1.5}$ km,
$N_{\rm H}=(0.72\pm0.02)\times10^{22}$ $\rm cm^{-2}$).

\subsubsection{Pulse Phase Spectroscopy with the 3$BB$ model}
\label{subsub:ppsbb}

\begin{figure}
\begin{center}
\includegraphics[angle=-90,scale=.40]{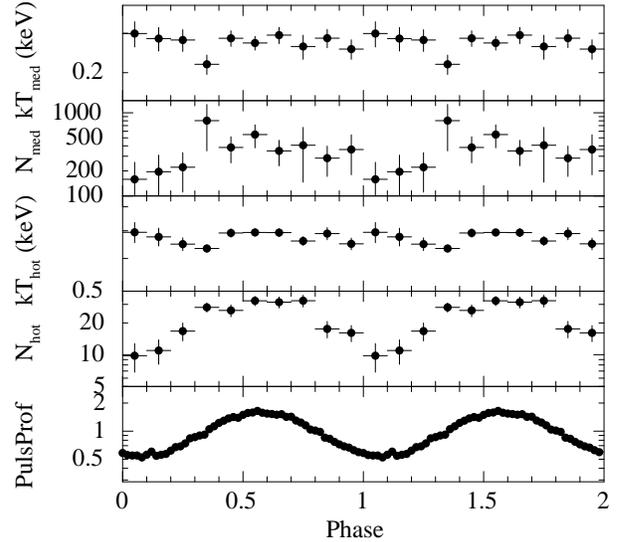}
\caption{Phase evolution of $BB_{\rm med}$ and $BB_{\rm hot}$ for the September 2003 observation.
The two temperatures remained constant to within the uncertainty, while the normalization changed; the ratio $\Delta$ between $N_{\rm max}$ and $N_{\rm min}$ is $\Delta N_{\rm hot,med}= N_{\rm max}/N_{\rm min}\sim3.3$ and
both peak at the same pulse phase ($\sim0.55$).}
\label{figure:ppsset03}
\end{center}
\end{figure}

\begin{figure}
\begin{center}
\includegraphics[angle=-90,scale=.30]{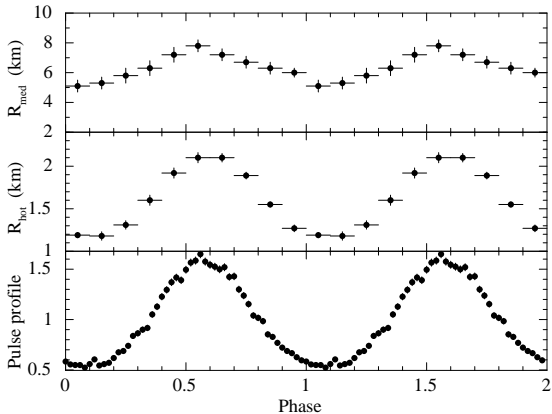}
\caption{$R_{\rm med}$ and $R_{\rm hot}$ as a function of rotation phase for the September
2003 observation. The temperatures $kT_{\rm med}$ and $kT_{\rm hot}$ are held fixed at the phase
average value listed in Table \ref{tab:3bbtot}, while the normalization
constant (which is related to the radius) is left free to vary.} 
\label{figure:ppsset03norm}
\end{center}
\end{figure}

\begin{figure}
\begin{center}
\includegraphics[angle=-90,scale=.30]{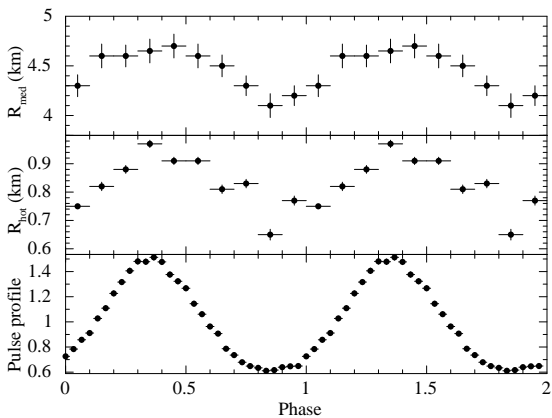}
\caption{$R_{\rm med}$ and $R_{\rm hot}$ as a function of rotation phase for the September
2004 observation. The temperatures $kT_{\rm med}$ and $kT_{\rm hot}$ are held fixed at the phase$-$averaged value listed in Table \ref{tab:3bbtot}, while the normalization constant (which is related to the radius) is set free to vary.} 
\label{figure:ppsset04norm}
\end{center}
\end{figure}

In the following we present the results from two representative cases:
the September 2003 and September 2004 observations. The PPS analysis
of the $BB$ components after the latter pointing was hampered by poor
statistics. All parameters were left free to vary except for
$kT_{\rm cold}$, $R_{\rm cold}$ and $N_{\rm H}$ that were frozen at the values
reported in Table \ref{tab:3bbtot}. We found that the temperature of
the medium and hot $BB$s were nearly constant through the whole pulse
cycle, whereas the normalization/emitting$-$area were clearly
variable (see Figure\,\ref{figure:ppsset03}). In order to better study
these variations, we fixed the $BB$ temperatures at their
phase$-$averaged values, leaving only the normalizations ($N$) of the 
spectra
free to vary\footnote{In the "blackbodyrad" model the normalization is
$N=R^2_{\rm km}/D^2_{10}$, with $D_{\rm 10}=0.35$ the distance to the source in
units of 10 kpc.}. These were then converted into the radii of the $BB$
components ($R_{\rm BB}=\sqrt{N}\times D_{10}$, assuming a source 
distance
of 3.5\,kpc). Figures\,\ref{figure:ppsset03norm},
\ref{figure:ppsset04norm} and Table \ref{tab:ppsset03norm} show our
results. In particular, in September 2003 the ratio $\Delta=R_{\rm max}/R_{\rm min}$ for each component was $\Delta
R_{\rm hot}=1.8\pm0.1$ and $\Delta R_{\rm med}=1.5\pm 0.2$, while in September 2004 $\Delta
R_{\rm hot}=1.3\pm 0.1$, $\Delta R_{\rm med}=1.2\pm 0.1$. In
both cases, the modulation of the radii ($R$) with phase shows only
one peak for pulse cycle. Moreover, they appear to be phase$-$aligned
with each other and with the peak of the total pulse profile. This
suggests that the two $BB$ regions must be relatively close to each
other and likely connected, otherwise a phase lag/shift would
naturally be expected. The $R$$-$variation amplitude as a function of
phase is more pronounced at higher energies, in agreement with the
timing properties of this pulsar, where the pulsed fraction is larger
at higher energies.\\
\begin{table}
\caption{$R_{\rm med}$ and $R_{\rm hot}$ as a function of pulse phase for the September
2003 and September 2004 observations. Temperatures are kept fixed at the value listed in Table
\ref{tab:3bbtot}. $1\sigma$ confidence level uncertainties are given. The
corresponding $\chi^2_{\rm red}$ are 0.92 (for 852 d.o.f.) and 1.01 (1132
d.o.f.) for September 2003 and 2004, respectively.}
\tiny
\begin{center}
\begin{tabular}{ccccc}
\hline \hline
          & Sept 03& Sept 03 & Sep 04 & Sep 04\\
Phase bin & $R_{\rm med}$  & $R_{\rm hot}$ & $R_{\rm med}$  & $R_{\rm hot}$ \\
          &   km       &   km    &    km       &   km        \\
\hline \hline 
0.0$-$0.1 & $ 5.1  \pm 0.4   $   & $ 1.19   \pm0.03   $  &  $4.3 \pm0.1   $   & $ 
0.750    \pm 0.001   $      \\
0.1$-$0.2&  $ 5.3  \pm 0.4   $   &  $1.18   \pm0.05   $  &  $ 4.6 \pm0.1   $   & $
 0.82    \pm  0.02  $       \\
0.2$-$0.3&  $ 5.8  \pm 0.5   $  &   $1.31   \pm0.05    $ &  $ 4.6 \pm0.1   $  &  
$ 0.88    \pm  0.01  $         \\
0.3$-$0.4&  $ 6.3  \pm 0.5   $  &   $ 1.60  \pm0.06   $  &  $ 4.6 \pm0.1   $  &  
$ 0.97    \pm  0.01  $      \\
0.4$-$0.5&  $ 7.2  \pm 0.5   $  &  $1.92    \pm0.06   $  &  $ 4.7 \pm0.1   $  &  
$ 0.91     \pm 0.01   $ \\
0.5$-$0.6&  $ 7.8  \pm 0.4   $   & $2.10    \pm0.06    $ &  $ 4.6 \pm0.1   $   & $
 0.91     \pm 0.01   $    \\
0.6$-$0.7&  $ 7.2  \pm 0.4   $   & $ 2.10   \pm0.05   $  &  $ 4.5 \pm0.1   $   & 
$ 0.81    \pm  0.02  $   \\
0.7$-$0.8&  $ 6.7  \pm 0.4   $  &   $1.89   \pm0.04   $  &   $4.3 \pm0.1   $  &  $
 0.83    \pm  0.02 $    \\
0.8$-$0.9&  $ 6.3  \pm 0.4   $  &   $ 1.55  \pm0.03   $  &   $4.1 \pm0.1   $  &   
$0.65     \pm 0.02   $    \\
0.9$-$0.1&  $ 6.0  \pm 0.3   $  &   $1.27   \pm0.04   $  &   $4.2 \pm0.1   $  &  
$ 0.77    \pm  0.02  $         \\
\hline \hline          
\end{tabular}
\label{tab:ppsset03norm}         
\end{center}
\end{table}
\label{subs:ppsbb}

\subsubsection{Pulse Phase Spectroscopy of the $\sim$1.1\,keV edge}
\label{subs:ppsedge}

\begin{figure}
\begin{center}
\includegraphics[angle=-90,scale=.35]{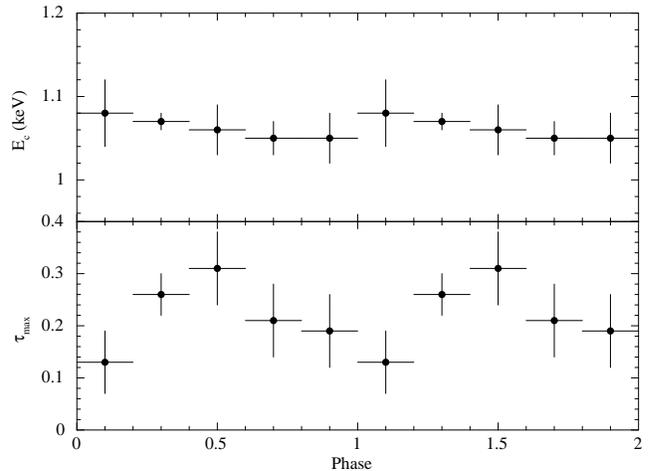}
\caption{Evolution of the edge parameters $E_{\rm c}$, and $\tau_{\rm max}$ with 
pulse phase for the September
2005 observation ($1\sigma$ confidence level uncertainties are reported).}  
\label{figure:edgppsset05}
\end{center} 
\end{figure}

A similar analysis was carried out for the narrow spectral feature detected in the
spectra from the March 2005 observation onwards. Given the relatively small number of photons, we reduced the
number of phase intervals to five and kept the spectral parameters 
of the coldest $BB$ fixed (3$BB$$+$edge model value). In Figure\,\ref{figure:edgppsset05} and Table \ref{tab:edgppsset05} we report the result for the September 2005 observation, when there was a possible indication that 
the component evolved with phase. Although the value of $\tau_{\rm max}$ is compatible with being constant ($\chi^2=6.57$ with 4 d.o.f.), 
we note that it varies from a minimum of $0.13\pm0.06$ to a maximum of $0.31\pm0.07$ in a smooth way, which we tried to model with a simple sinusoidal function. An F$-$Test for the addition of the sinusoid gave just a marginal detection ($\sim2.4\sigma$), hence no claim can be made about its actual presence. However this possible modulation is worth further investigation with deeper observations.

\begin{table}
\caption{Evolution of the spectral parameters for the edge component as a
function of phase for the September 2005 observation (PN data). $1\sigma$
confidence level are reported. The five spectra are fitted together and
resulting in a $\chi^2_{\rm red}$ of 1.01 (for 406 d.o.f.).}     
\begin{center}
\begin{tabular}{ccc} 
\hline \hline
Phase bin & $E$  & $\tau_{\rm max}$ \\
     &     keV    &        \\
\hline \hline 
0.0$-$0.2 &1.08   $\pm0.04      $   &  $    0.13    \pm  0.06 $  \\
0.2$-$0.4 &1.07   $\pm0.01     $   &     $  0.26    \pm  0.04 $  \\
0.4$-$0.6 &1.06   $\pm0.03    $  &       $  0.31    \pm  0.07 $   \\
0.6$-$0.8 &1.05   $\pm0.02    $  &       $  0.21    \pm  0.07 $  \\
0.8$-$1.0 &1.05   $\pm0.03     $ &       $  0.13    \pm  0.07 $  \\
\hline \hline          
\end{tabular} 
\label{tab:edgppsset05}
\end{center} 
\end{table}
\section{Discussion}
\label{diss}
The spectral and temporal information obtained from the nine \XMM\
observations of the Transient Anomalous X$-$ray Pulsar \xte\ collected
in 2003$-$2007 allowed us to study to an unprecedented level of detail the source
behavior during the outburst. As discussed below, our results shed some
light on several issues concerning the mechanism powering the emission
during the active period. During four years of monitoring, the X$-$ray
flux of \xte\ continued to decrease following an almost exponential
decay. In September 2007 the source nearly reached its quiescent emission level
as recorded by \textit{ROSAT} in 1992. In the following we summarize the most relevant findings that we obtained from
the \XMM\ dataset.
\subsection{The continuum spectral component}
We found that the previously proposed 2$BB$ model for the source
spectrum during the outburst fails to account for the time evolution of the hot$-$temperature spectral components and for the PF flattening (see \S\ref{sub:2bb-3bb}). Similar concerns have been expressed already by Israel et al (2007b) and Perna \& Gotthelf (2008), but see e.g. G{\"u}ver et
al. (2007) for a different interpretation. For these reasons we included
a third softer thermal component which, as we have shown in
the previous sections, much improves the spectral fits and also
removes the inconsistency that appears in the 2$BB$ model when the
evolution of the PF is considered. The
temperature and radius of this additional $BB$ turn out to be the same as
those inferred from the \xte\ \textit{ROSAT} spectra 
serendipitously collected since 1992, when the source was in
quiescence.\\ The additional $BB$ component is compatible with being
emitted from the whole NS surface, and appears to be unaffected by the
outburst. Therefore, its nearly constant flux can be taken as
representative for the minimum level of emission from the source. It also provides the key to understanding the previously
unexplained PF flattening. We emphasize that the 3$BB$ model discussed here should be regarded as a
crude, albeit convenient, description of a scenario in which other
effects may come into play (see below). Nevertheless, it has the
advantage of being independent of the, often poorly known,
details of the atmosphere/magnetosphere of the neutron star. As
such, it provides a first estimate of some key physical parameters,
like the size and temperatures of the emitting region(s), without relying
on any assumptions about the field strength or geometry. 
Intriguingly, our analysis reveals that only the size of the hot/warm regions varied during the
outburst, showing an almost steady decrease, while the temperatures
remained nearly constant.\\ 
Although present data do not allow to tightly constrain the shape and
relative position of the hot and medium temperature regions on the
star, a simple model can be used to gain some insight on the geometry
of the source. We assume that emission comes from two concentric
zones: an inner, hot cap, and an outer, warm corona, outside of which
is the colder surface of the star at $T_{\rm cold} = 0.160$ 
keV\footnote{The radii of these regions
are taken from table \ref{tab:ragtempPBB}.}, a picture very similar to
that adopted by Israel et al. (2007b). For a NS of 1.4 $M_\odot$ and typical NS radii, we computed the PFs after applying the proper
relativistic corrections. Since the angular (semi)aperture of the two
zones follows from the values of the blackbody radii, and their
temperatures are just $kT_{\rm hot}$ and $kT_{\rm med}$, the only free
parameter is the angle between the diameter through the cap center and
the rotation axis, i.e. the cap's colatitude. The observed
spectrum and the lightcurves also depend on the angle between the
line of sight (LOS) and the rotation axis. Without performing any
formal fit, we simply tried various combinations of these angles, and
we found that there is reasonable agreement between PF data and model 
at all epochs for
values which are consistent with the range determined by the detailed
analysis of Perna \& Gotthelf (2008; see also Kramer et al. 2007 for constraints on the pulsar geometry through radio polarimetry). This is
not unexpected since the spectrum in the first 4 epochs, which Perna \&
Gotthelf analyzed, is not much affected by the emission from 
the coldest part of the star surface.\\ 
We note that our analysis based on the 3$BB$ model suggests
that the coldest $BB$ component, accounting for the emission from the
whole surface, has a low pulsed fraction, $10\%\,\pm1\%$. If our
model is correct, this prediction can be checked once the
source returns to the quiescent state. We also note that this value is
similar to that found in X$-$ray Dim Isolated Neutron Stars (XDINSs),
where it is believed that the (purely) thermal emission comes from the
cooling NS surface (e.g. Haberl 2007). Although magnetars as a class
are probably far from being passive coolers, this similarity makes a
case for our interpretation of the cold $BB$ component as the quiescent
emission from the NS surface, worth being pursued in future studies.\\
On the other hand, the narrower pulse profile and larger pulsed
fraction at increasing energies seems reminiscent of what was found
for other AXPs with \RXTE\ and $INTEGRAL$ in the energy band above 10
keV. Indeed, the narrowing of the peak is coincident with the
emergence of a hard power$-$law component extending from 10$-$20\,keV up
to 200\,keV at least (Kuiper et al. 2004). The origin of this
component is most likely magnetospheric. The marginal detection,
during the first three \XMM\ pointings, of a possible hard
power$-$law tail extending above 10\,keV, corroborates this
reasoning. However we could not study in more detail the power$-$law
tail, due to insufficient statistics.\\ We performed also a
preliminary test with a different model, a simple Comptonization
model available in $XSPEC$ ($CompTT$), but this gives a worse fit for the
data with respect to the 3$BB$ model. More advanced RCS models in which the
optical depth is provided by currents flowing in a twisted
magnetosphere (Lyutikov and Gavriil 2006; Fernandez \& Thompson 2007;
Nobili, Turolla \& Zane 2008a, b) appear, on the other hand, promising
in explaining the pulse profiles, since the particle density changes
with the magnetic colatitude, increasing as one moves from the
magnetic pole towards the equator. Such a distribution naturally
introduces a pulsed fraction even in the case in which the surface
temperature is homogeneous, as recently shown on the basis of
Montecarlo simulations by Nobili et al. (2008a) and Pavan et al. (in
preparation). A first attempt to systematically apply RCS to all AXPs,
including \xte\, has been reported by Rea et al.~(2008). These
authors found that the outburst of this source may result from 
heating of the NS surface, which slowly cools on a timescale of
months/years, while the magnetospheric properties show only small
variation during the outburst decay.\\

\subsection{The narrow feature at $\sim1\,\rm keV$}
\label{section:narrow}

Within the framework of the magnetar model, a natural interpretation
for
the absorption$-$like feature which is significantly detected in the PN and MOS
spectra is that it is due to a proton cyclotron line.
The observation of such a feature would directly probe the magnetic field strength of the AXP, since the line energy is
proportional to the field strength:
\begin{equation}
E_{\rm cyc}= 0.63 (1+z)^{-1} \left(\frac{B}{10^{14}\, G}\right)\,\rm keV
\label{eq:cyclprot}
\end{equation}
where $(1+z)^{-1}=(1-2GM/Rc^2)^{\frac{1}{2}}\simeq0.8$ is the
gravitational redshift at the neutron star surface. Here we assumed
$M=1.4M_{\rm \odot}$ and $R=10$ km for the star mass and radius. Despite a few earlier
claims (Ibrahim et al. 2002; Rea et al. 2003),
unambiguous evidence of the presence of absorption lines in the spectra of
magnetars has not yet been obtained.\\
If the edge detected in the \xte\ spectra is a proton cyclotron
feature, when taken face value its energy implies
$2.1\times10^{14}$ G $\leq
B_{\rm prot}$$\leq 2.6\times10^{14}$ G.
On the other hand, the assumption of a constant field breaks down if the
line originates from a relatively large region on the neutron
star surface/magnetosphere. For instance, Zane et al. (2001) estimated that, even for a simple dipolar field,
the fact that $B$ changes in both
magnitude and direction will produce a
broadening of a feature which is emitted by the whole surface (typically
by 10\%$-$20\%) and a shift of the line centroid toward
lower energies by 20\%$-$30\% with respect to the prediction based on
eq~(\ref{eq:cyclprot}).

Similar absorption features are also observed in the spectra of XDINSs 
(Haberl 2007) and are typically associated with
proton cyclotron and/or bound$-$free, bound$-$bound transitions in H, H$-$like
and He$-$like atoms in the presence of relatively high magnetic
fields $B\approx 5 \times 10^{13}$--$10^{14}$~G
(e.g. van Kerkwijk \& Kaplan 2007; Ho et al., 2003; Pavlov \& Bezchastnov, 2005). At such large
field strengths, exotic molecules might also contribute to line formation
(Turbiner et al.~2007, Turbiner \& Lopez$-$Vieyra, 2006). For XDINSs, all 
the above mentioned scenarios provide similar values of $B$, which turns out to be 
in agreement with those derived from the spin$-$down rate (e.g. Kaplan 2008).
A similar absorption feature has been discovered in the spectrum of the Rotating RAdio Transient
(RRAT) detected at X$-$ray energies, J1819$-$1458
(McLaughlin et al. 2007). The X$-$ray spectrum of RRAT J1819$-$1458 is well
fit by an absorbed blackbody with $kT=0.14$~keV with the addition
of an absorption feature at $\sim 1$ keV, which, when interpreted either
as a proton cyclotron line or as an atomic transition, yields a magnetic field of $5\times10^{13}$\,G,
again in rough agreement with the spin$-$down measure (McLaughlin et al. 2007).
Also in the case of \xte, the magnetic field value inferred by using eq.(~\ref{eq:cyclprot})
appears to be in very good agreement with that obtained
through the spin$-$down measurement:
$2.2\times10^{14}\,{\rm G}\leq B_{\rm dip}\leq3.1\times10^{14}\,\rm G$.
It is interesting to note that a
similar value, $B=(2.72\pm0.03)\times10^{14}$\,G was obtained by G\"uver et al. (2007) based on the
September 2003 $-$ March 2006 \XMM\ spectrum of \xte. It is worth
emphasizing that
the spectral model used by G\"uver et al. (2007), has been
specifically developed for passively cooling NS and magnetic field
stronger than
$5\times10^{13}$\,G and is, therefore, rather different from the 3$BB$
model adopted here.

A different possibility is that the line is due to the presence of Iron
in proximity of the star surface. In particular, L shell electronic
transitions of Iron ions XXII, XXIII, XXIV, have energies between 1.05
and 1.17 keV. However this requires that the line absorbing
region is permeated by a relatively low magnetic field. Future longer observations, with much higher statistics, might help to better understand the nature of this spectral feature.

\subsection{Flux evolution}
\begin{figure}
\begin{center}
\includegraphics[angle=-90,scale=.34]{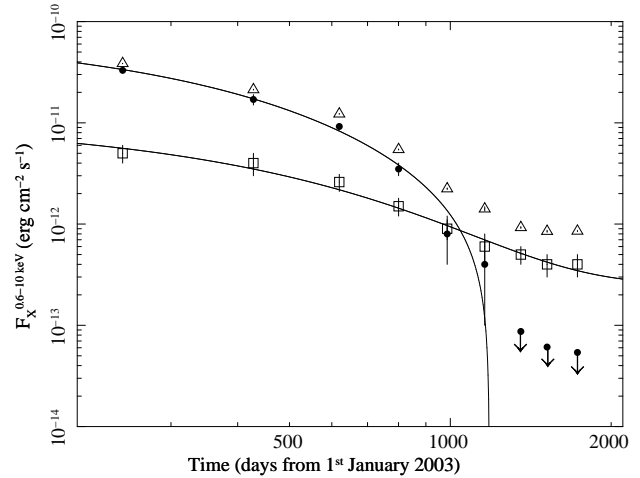}
\caption{Evolution of the $0.6-10$ keV flux (as measured with the EPIC/PN 
camera on board \XMM\ of \xte\ as a function
of time, for the $BB_{\rm med}$ (squares), the $BB_{\rm hot}$ (circles), and the sum of
the 3$BB$s (triangles). The solid line represents the best fit obtained for the
$BB_{\rm med}$ and 3$BB$ evolution by using a model consisting of an exponential decay plus a
constant.}
\label{figure:xtevssgr}
\end{center}
\end{figure}

During approximately four years of \XMM\ monitoring, the X$-$ray flux of
\xte\ continued to decrease, and is presently $\sim$15\%$-$20\% above
the quiescent level (as determined by \textit{ROSAT}). In Figure
\ref{figure:xtevssgr} the evolution of the total X$-$ray flux in the
0.6$-$10\,keV band is shown (triangles), together with the flux
evolution of the two hotter $BB$s, $BB_{\rm med}$ (squares) and $BB_{\rm
hot}$ (circles). Notably, both the $BB_{\rm med}$ and $BB_{\rm hot}$
flux evolutions are well fit by an exponential decay plus a constant
($\chi^2=2$ for 5 d.o.f. and $\chi^2=5$ for 2 d.o.f.,
respectively). The characteristic times are $\tau=370\pm40$ days
and $\tau=250\pm10$ days for $BB_{\rm med}$ and $BB_{\rm hot}$,
respectively. This might hint towards a common physical process
responsible for the decay of the two $BB$ components, though on slightly
different timescales. A possible flattening in the $BB_{\rm med}$ flux
evolution, as suggested by the latest two/three flux measurements,
might imply that this component has already reached its quiescent
state (see discussion below).
\begin{figure*}[tbh]
\begin{center}
\includegraphics[angle=-90,scale=.55]{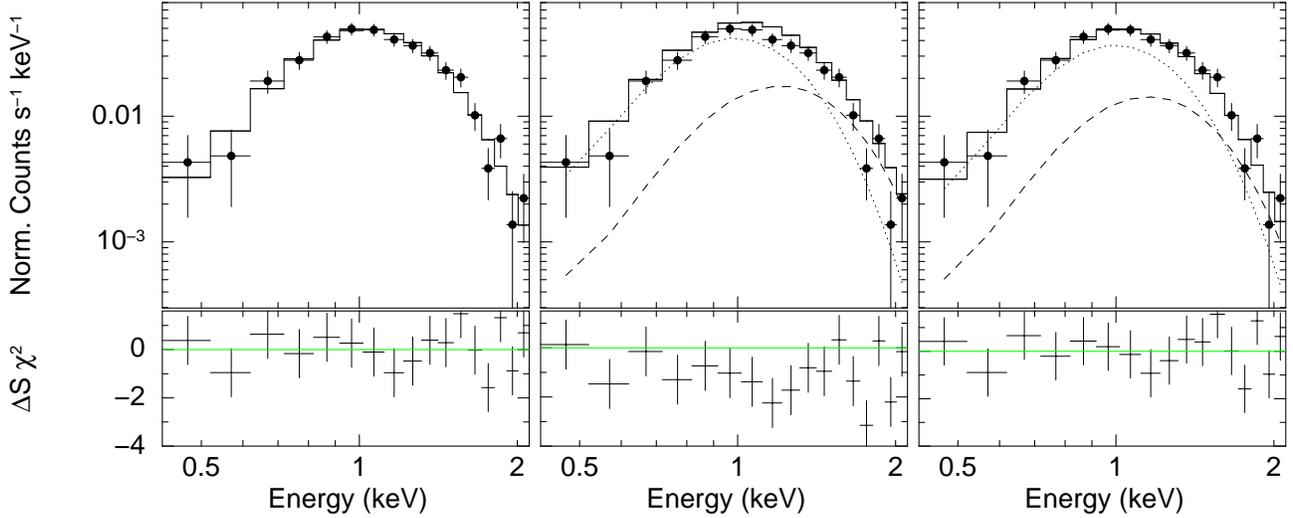}
\caption{\textit{ROSAT} PSPC spectrum of the pre$-$outburst quiescent state of
\xte\ (three PSPC observations were merged together to rely upon a higher S/N),
fitted with one $BB$ component (left panel), same \textit{ROSAT} spectrum with superimposed
the 2$BB$ model inferred from the latest three \XMM\ observations (since
September 2006 the $BB_{\rm hot}$ is not detected; no fit was performed; middle
panel), same as before but leaving free to vary the 2$BB$ model parameters (right
panel). The dotted and dash$-$stepped lines mark the $BB_{\rm cold}$ and $BB_{\rm med}$
components, respectively.
}
\label{figure:xmmros}
\end{center}
\end{figure*}

In order to further test this hypothesis we superimposed the average
spectral model, referred to the latest three \XMM\ observations
(September 2006 $-$ 2007, where only the $BB_{\rm cold}$ and $BB_{\rm med}$
components are detected), to the average \textit{ROSAT} spectrum
obtained by merging the three longest pointings (total effective
exposure of $\sim22$\,ks). This model is compared with the single $BB$
model used so far for the \textit{ROSAT} data. The result of this
test is shown in Figure\,\ref{figure:xmmros}. It is evident from the
first and second panel that the September 2006 $-$ 2007 \XMM\ model is
in agreement with the \textit{ROSAT} data in consideration of the fact
that no fit has been performed, suggesting that the source might be
already back to its quiescent state since March 2007. If correct, the
quiescent state of \xte\ could be characterized by the presence of two
$BB$s instead of one $BB$ as discussed so far. However, we emphasize that
the inclusion of the second $BB$ in the \textit{ROSAT} spectral fit is
formally not statistically required. In fact we reanalyzed the
\textit{ROSAT} data by using either a single $BB$ or a 2$BB$ model, and
in both cases we obtained a $\chi^2_{\rm red}$=0.9 (left and right
panels of Figure\,\ref{figure:xmmros}). The best fit parameters are:
($BB$) $N\rm _H$=$(0.63\pm0.05)\times10^{22}$cm$^{-2}$,
$kT=0.19\pm0.03$\,keV and $R<11$ km ($\chi^2=13$ for 14 d.o.f.),
(2$BB$) $N_{\rm_H}=(0.75\pm0.08)\times10^{22}$cm$^{-2}$, $kT_{\rm cold}=0.16
\pm0.03$\,keV and $R_{\rm cold}=16\pm5$ km,
kT$_{\rm med}=0.26\pm0.06$ keV and $R_{\rm med}<5$ km ($\chi^2=$11 for 12
d.o.f.). On the other hand, we
note that the \XMM\ model remains slightly above the \textit{ROSAT}
data mainly around 1\,keV, where the $BB_{\rm med}$ component is
maximum. This might suggest that the flux of the latter component
is still decaying. Clearly, a deeper and higher$-$statistics observation
of \xte\ at some later time might solve this issue.

\section{Conclusions}
In this paper we reported the detailed timing and spectral analysis of a
long$-$term (4 years) \XMM\ monitoring program aimed at unveiling the physical
processes responsible for the decaying phases of the \xte\ outburst.
The main results can be summarized as follows:
\begin{itemize}

\item We found that a spectral model with three blackbodies is in
much better agreement with the data than the previously used model
involving two blackbodies. Also, the 3$BB$ model solves several
ambiguities in the spectral evolution that were present in the 2$BB$
model.

\item The best spectral fit at the different epochs is obtained for three 
blackbodies plus an edge. 
The best fit spectral parameters determined with this model are: $kT_{\rm cold}\sim0.15$ keV and
$R_{\rm cold}\sim15$\,km. The latter feature is required starting from the March 2005 observation, where residuals with respect to the simple 3$BB$ model are clearly recognized.
The coldest $BB$ component temperature and emitting radius remain constant during the whole outburst and are the same as those of the single $BB$ component observed by \textit{ROSAT}, which is
likely emitted from the whole NS surface.
The two hotter and smaller regions ($\sim5$ and $\sim1$ km) evolve in size but, again, at
constant temperature. The emitting surface decreases in both cases and these components are,
therefore, likely responsible for the enhancement of the observed X$-$ray
flux during the outburst.

Since September 2006 the hottest component, $BB_{\rm hot}$, is no longer needed in
the fit and the 3$BB$ model evolves into a 2$BB$ model. At the same epoch,
the average pulsed fraction of the 5.54\,s modulation levels up suggesting that
the greatest part of the pulsed photons were produced in the $BB_{\rm hot}$
component.

\item During the first three \XMM\ observations (2003$-$2004) the
spectral fit residuals suggest the presence of an additional component
above 7$-$8\,keV, probably a hard tail, possibly similar to the
one detected in other AXPs (where it extends up to 200 keV). The
limited sensitivity of the EPIC cameras above 10\,keV prevented us
from performing a detailed analysis of this component.

\item By assuming that the feature around 1.1 keV is due to a proton cyclotron resonance,
we obtain a surface magnetic field value of
$2.1\times10^{14}\, G \leq B_{\rm prot}\leq2.6 \times10^{14}\, G$. This estimate
is in very good agreement with that obtained from the spin$-$down measure of
$1.6\times10^{14}\,{\rm G}\leq B_{\rm dip}\leq2.8\times10^{14}\,\rm G$.
We can not currently exclude that the absorption feature originates from L$-$shell transitions of Fe XXII, XXIII and XXIV.

\item The analysis of the pulsed fraction time evolution as a function of energy
shows an increase with energy, within individual observations, and a decrease
as a function of time, within the same energy interval. Most of the modulation is
ascribed to high$-$energy photons coming from the two hottest $BB$ emitting regions.

\item Pulse phase spectroscopy shows that emission from the two hotter $BB$s peaks at the same phase interval,
suggesting that they are emitted by close$-$by regions (e.g. two concentric zones).
\item The observed (0.6$-$10 keV) flux evolution of the $BB_{\rm med}$ and
$BB_{\rm hot}$ is well described by an exponential decay, with characteristic times of
$\tau=370\pm40$ days and $\tau=250\pm10$ days, respectively. This suggests that the same physical process
is responsible for the decay of the two thermal components, as already noted by Gotthelf \& Halpern (2005).
While, in a 2$BB$ model, the hot component shows similar time decay ($\tau\sim300$ days), the decay time of the
colder one is longer ($\sim900$ days). This is in agreement with the presence of a colder component, emitted by
the whole the star surface. 
\item A comparison between the latest three \XMM\ pointings and a re$-$analysis of the
\textit{ROSAT} quiescent spectrum reveals that the $BB_{\rm med}$ component might have
already reached its quiescent state.
\end{itemize}

\begin{acknowledgements}
This work is partially supported at OAR through Agenzia Spaziale Italiana (ASI),
Ministero dell'Istruzione, Universit\`a e Ricerca Scientifica e Tecnologica
(MIUR $-$ COFIN), and Istituto Nazionale di Astrofisica (INAF) grants. We
acknowledge financial contribution from contract ASI$-$INAF I/023/05/0 and AAE TH$-$058. Based on
observations obtained with \XMM, an ESA science mission with instruments
and contributions directly funded by ESA Member States and NASA. This research
has made use of data obtained through the High Energy Astrophysics Science
Archive Research Center Online Service, provided by the NASA/Goddard Space
Flight Center. SZ acknowledges support from a STFC (ex$-$PPARC) AF. DG
acknowledges financial support from the French Space Agency (CNES).

\end{acknowledgements}

\end{document}